\documentclass[aps,prl,floatfix,showpacs,twocolumn,longbibliography,superscriptaddress]{revtex4-1}
\usepackage{ulem}
\usepackage{bm}
\usepackage{color}
\usepackage{amssymb}
\usepackage{amsmath}
\usepackage{graphicx}
\usepackage{amsfonts}
\usepackage{slashed}
\usepackage{pstricks}
\usepackage{float}
\usepackage{hyperref}
\usepackage{array}
\allowdisplaybreaks
\usepackage{dcolumn}
\usepackage{epsf}
\usepackage{tabularx}
\usepackage{soul}
\usepackage{wrapfig}

\newcommand{\bs}[1]{\boldsymbol{{#1}}}

\begin{document}
\title{Dilute neutron star matter from neural-network quantum states}

\author{Bryce Fore}
\affiliation{Physics Division, Argonne National Laboratory, Argonne, IL 60439, USA}

\author{Jane M. Kim}
\affiliation{Department of Physics and Astronomy and Facility for Rare Isotope Beams, Michigan State University, East Lansing, MI 48824, USA}

\author{Giuseppe Carleo}
\affiliation{Institute of Physics, École Polytechnique Fédérale de Lausanne (EPFL), CH-1015 Lausanne, Switzerland}

\author{Morten Hjorth-Jensen}
\affiliation{Department of Physics and Astronomy and Facility for Rare Isotope Beams, Michigan State University, East Lansing, MI 48824, USA}
\affiliation{Department of Physics and Center for Computing in Science Education, University of Oslo, N-0316 Oslo, Norway}

\author{Alessandro Lovato}
\affiliation{Physics Division, Argonne National Laboratory, Argonne, IL 60439}
\affiliation{Computational Science Division, Argonne National Laboratory, Argonne, IL 60439}
\affiliation{INFN-TIFPA Trento Institute of Fundamental Physics and Applications, 38123 Trento, Italy}

\date{\today}
\begin{abstract}
Low-density neutron matter is characterized by fascinating emergent quantum phenomena, such as the formation of Cooper pairs and the onset of superfluidity. We model this density regime by capitalizing on the expressivity of the hidden-nucleon neural-network quantum states combined with variational Monte Carlo and stochastic reconfiguration techniques. Our approach is competitive with the auxiliary-field diffusion Monte Carlo method at a fraction of the computational cost. Using a leading-order pionless effective field theory Hamiltonian, we compute the energy per particle of infinite neutron matter and compare it with those obtained from highly realistic interactions. In addition, a comparison between the spin-singlet and triplet two-body distribution functions indicates the emergence pairing in the $^1S_0$ channel.  
\end{abstract} 
%
%
\maketitle

\paragraph{Introduction.} Multi-messenger astronomy has opened new windows into the state of matter at densities and isospin asymmetries that cannot be directly probed by terrestrial experiments~\cite{LIGOScientific:2017vwq,Abbott:2017,Sabatucci:2020xwt,senger2021}. Concurrently, nuclear many-body theory has made considerable progress in computing the nucleonic-matter equation of state at densities corresponding to the inner core of neutron stars starting from realistic Hamiltonians~\cite{Drischler:2016djf, Piarulli:2019pfq, Lonardoni:2019ypg, Jiang:2020the, Sammarruca:2021bpn, Sammarruca:2021bpn,Heiselberg2000}. Comparisons between theoretical predictions and astrophysical observation pose stringent constraints on models of nuclear dynamics, particularly three-nucleon forces~\cite{Sabatucci:2022qyi}. 

In this work, we focus on lower densities, $\rho \lesssim 0.04$ fm$^{-3}$, which are relevant to the phenomenology of the stellar inner crust and outer core. In this region, both conditions for superfluidity --- strong Fermi degeneracy and an attractive interaction between neutron pairs in the $^1S_0$ channel --- are believed to be met~\cite{Sedrakian:2006,Benhar:2017mof,Dean2003}. In addition to lowering the system's energy, the formation of Cooper pairs plays a critical role in neutrino emission~\cite{Yakovlev:2004iq,Page:2010aw}, and the phenomenology of glitches~\cite{Monrozeau:2007xu}. Pairing is also relevant in modeling neutron-rich nuclei, which are the subject of intense experimental activities~\cite{Nowacki:2021fjw}.

Quantum Monte Carlo approaches~\cite{Carlson:2014vla}, and in particular the auxiliary-field diffusion Monte Carlo (AFDMC) method~\cite{Schmidt:1999lik} have been extensively applied to accurately compute neutron-matter properties~\cite{Lonardoni:2019ypg,Piarulli:2019pfq,Lovato:2022apd}. In the low-density regime, AFDMC calculations have convincingly shown a depletion of the superfluid gap with respect to the Bardeen–Cooper–Schrieffer theory~\cite{Gandolfi:2008id,Gandolfi:2022dlx}. However, because of the fermion sign problem, AFDMC predictions depend upon the starting variational wave function. For instance, the superfluid phase must be assumed {\it a priori} by using pfaffian wave functions~\cite{Bajdich:2006zz}. 

Neural-network quantum states~\cite{carleo_solving_2017} (NQS) have gained popularity in solving the Schr\"odinger equation of atomic nuclei both in real space~\cite{Keeble:2019bkv,Adams:2020aax,Gnech:2021wfn,Lovato:2022tjh,Yang:2022esu} and in the occupation-number formalism~\cite{Rigo:2022ces}. In this work, we introduce a periodic NQS suitable to model both the normal and superfluid phases of neutron matter. The ansatz is based on the ``hidden-nucleon'' architecture, which can model the ground-state wave functions of nuclei up to $^{16}$O with high accuracy~\cite{Lovato:2022tjh}. Inspired by chemistry applications~\cite{Hermann:2019,Pfau:2019}, we further improve the expressivity of the hidden-nucleon NQS using generalized backflow correlations, which generalize both the pfaffian and the spin-dependent backflow of Ref.~\cite{Brualla:2003gw}. 

Our model of nuclear dynamics is the leading-order pionless effective field theory ($\slashed{\pi}$EFT) Hamiltonian of Ref.~\cite{Schiavilla:2021dun}, which qualitatively reproduces the binding energies of nuclei with up to $A=90$ nucleons. Arguments based on the expansion around the unitary limit~\cite{Konig:2016utl}, and Brueckner-Hartree-Fock calculations of infinite nuclear matter~\cite{Kievsky:2018xsl}, indicate that $\slashed{\pi}$EFT should provide accurate energies of dilute neutron matter. We test this hypothesis by comparing the $\slashed{\pi}$EFT energy per particle against the sophisticated Argonne $v_{18}$~\cite{Wiringa:1994wb} plus Urbana IX~\cite{Pudliner:1995wk} (AV18+UIX) Hamiltonian used in the Akmal-Pandharipande-Ravenhall (APR)~\cite{Akmal:1998cf} equation of state. 

To better quantify the role of dynamical correlations, we evaluate the two-body spatial distribution functions, separating the spin-triplet and spin-singlet channels. We analyze the self-emergence of pairing correlations, not explicitly included in the NQS ansatz, as a function of neutron-matter density.  

\paragraph{Method.} We model the interactions among neutrons through the leading-order $\slashed{\pi}$EFT Hamiltonian ``o'' of Ref.~\cite{Schiavilla:2021dun}. The two-body contact potential is designed to reproduce the $np$ scattering lengths and effective ranges in the $S/T = 0/1$ and $1/0$ channels. Thus, it yields a neutron-neutron scattering length of $a_{nn}=-22.5$ fm, slightly larger than the experimental value of $-18.9(4)$ fm, see \cite{Machleidt2011} and references therein, while the effective range is well reproduced. The Hamiltonian also contains a repulsive three-body force that ensures the stability of nuclei. 

We approximate the ground-state solution of the nuclear many-body problem with an NQS ansatz that belongs to the hidden-fermion family~\cite{Moreno2022}, recently generalized to continuum Hilbert spaces and applied to atomic nuclei in Ref.~\cite{Lovato:2022tjh}. In addition to the visible spatial and spin coordinates of the $A$ neutrons, $R=\{\mathbf{r}_1\dots \mathbf{r}_A\}$ and $S=\{s^z_1\dots s^z_A\}$, the Hilbert space contains fictitious $A_h$ hidden-nucleon degrees of freedom. In this work we use $A_h=A=14$ so that the system is as flexible as possible, but in practice we have also found using as few as $8$ hidden nucleons gives very similar results. 
The wave function can be conveniently expressed in a block matrix form as
\begin{equation}
    \Psi_{HN}(R, S) \equiv \text{det}\left[ 
    \begin{matrix}
        \phi_v(R, S) & \phi_v(R_h, S_h)\\
        \chi_h(R, S) & \chi_h(R_h, S_h)
    \end{matrix}
    \right]\,.
\end{equation}
As in Ref.~\cite{Lovato:2022tjh}, $\phi_v(R,S)$ is the $A \times A$ matrix representing visible single-particle orbitals computed on the visible coordinates while the $A_h \times A_h$ matrix $\chi_h(R_h,S_h)$ yields the amplitudes of hidden orbitals evaluated on the coordinates of the $A_h$ hidden nucleons. Finally, $\chi_h(R,S)$ and $\phi_v(R_h,S_h)$ are $A_h \times A$ and $A \times A_h$ matrices giving the amplitudes of hidden orbitals on visible coordinates and visible orbitals on hidden coordinates, respectively. All the above matrices are expressed in terms of deep neural networks with differentiable activation functions --- see Ref.~\cite{Lovato:2022tjh} for additional details. To respect the Pauli principle, the coordinates of the hidden nucleons must be permutation-invariant functions of the visible ones. We enforce this symmetry by using a Deep-Sets architecture~\cite{Zaheer:2017,Wagstaff:2019} with {\it logsumexp} pooling. 

Inspired by the success of quantum-chemistry NQS~\cite{Hermann:2019,Pfau:2019}, we augment the flexibility of the ansatz by performing a generalized backflow transformation to the visible coordinates entering the upper-left block of the hidden-nucleon matrix: $\phi_v(R,S)\to \phi_v(\tilde{R},\tilde{S})$. We use the Deep-Sets architecture again to enforce fermion anti-symmetry
\begin{equation}
(\tilde{\mathbf{r}}_i, \tilde{s}^z_i) = \rho_{\rm bf}\Big(\mathbf{r}_i, s^z_i, \log \Big( \sum_j{\exp(\phi_{\rm bf}(\mathbf{r}_j, s^z_j)}\Big)\Big)\,.
\end{equation}
To further augment the expressivity, each visible single-particle orbital uses its own $\rho_{\rm bf}$ and $\phi_{\rm bf}$ neural networks. 

We simulate infinite neutron matter using $14$ particles in a box with periodic boundary conditions. Following Ref.~\cite{Pescia:2021kxb}, the latter are imposed by mapping the spatial coordinates onto periodic functions by 
\begin{equation}
    \bs{r}_i \rightarrow \left(\sin\left(\frac{2\pi \bs{r}_i}{L}\right), \cos\left(\frac{2\pi \bs{r}_i}{L}\right)\right)\,
\end{equation}
which ensures the wave function is continuous and differentiable at the box boundary. Here $L$ is the size of the simulation periodic box, and the $\sin$ and $\cos$ functions are applied element-wise to $\bs{r}_i$. Finite-size effects due to the tail corrections of two- and three-body potentials are accounted for by summing the contributions given by neighboring cells to the simulation box~\cite{Sarsa:2003zu}. 
\begin{figure}[b]
    \includegraphics[width=\columnwidth]{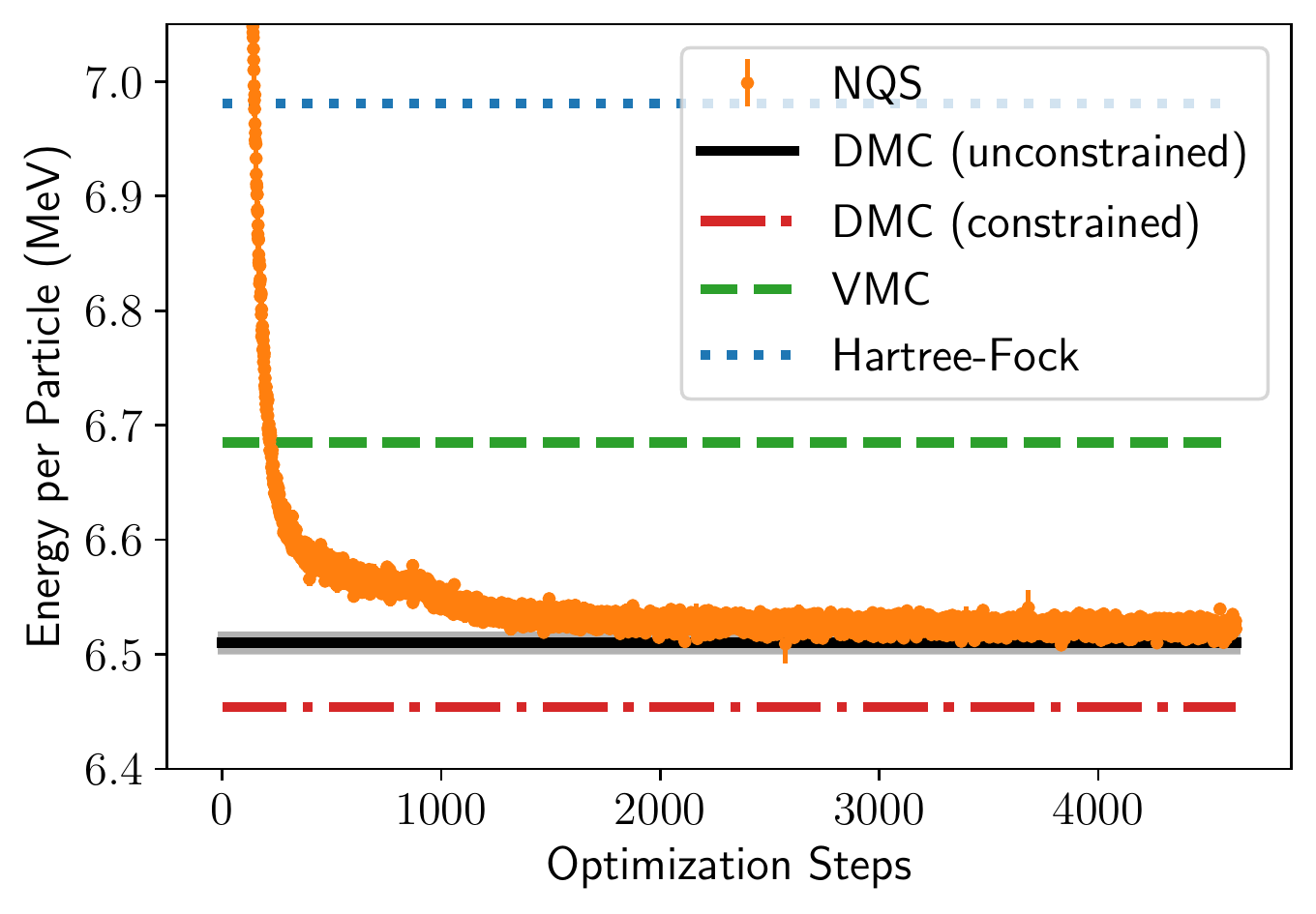}
\caption{NQS training data in neutron matter at $\rho = 0.04$ fm$^{-3}$ (data points) compared with Hartree-Fock (dotted line), conventional VMC (dashed line), constrained-path ADMC (dash-dotted line) and unconstrained-path ADMC results (solid line).}
\label{fig:training}
\end{figure}

Evaluating the expectation values of quantum mechanical operators, including the Hamiltonian, requires carrying out multi-dimensional integration over the spatial and spin coordinates of the neutrons. To this aim, we exploit Monte Carlo quadrature and sample $R$ and $S$ from $|\Psi_{HN}(R,S)|^2$ using the Metropolis-Hastings algorithm~\cite{Metropolis:1953} --- additional details can be found in the supplemental material of Ref.~\cite{Adams:2020aax}. The best variational parameters defining the NQS are found by minimizing the system's energy, which we carry out using the R(oot)M(ean)S(quared)Prop(agation)-enhanced version of the stochastic-reconfiguration optimization method introduced in Ref.~\cite{Lovato:2022tjh}. 

\paragraph{Results and discussion.} We first benchmark the expressivity of the hidden-nucleon NQS for periodic systems by comparing the energy per particle of infinite neutron matter against ``conventional'' variational Monte Carlo (VMC), and both constrained-path and AFDMC results. The variational wave function used in state-of-the-art neutron-matter studies, see for example \cite{Lonardoni:2019ypg,Lovato:2022apd}, contains a spin-independent Jastrow factor that multiplies a Slater determinant augmented by spin-dependent backflow correlations. The constrained-path approximation, commonly employed to alleviate the AFDMC fermion-sign problem~\cite{Carlson:2014vla}, brings about a bias in the ground-state energy estimate~\cite{Piarulli:2019pfq,Lovato:2022apd}. Exact results can be obtained by performing unconstrained propagations, but the statistical error grows exponentially with the imaginary time. 

As shown in Fig.~\ref{fig:training} for $\rho = 0.04$ fm$^{-3}$, after $\simeq 2000$ stochastic-reconfiguration steps, the NQS ansatz converges to the {\it virtually exact} unconstrained AFDMC energy, using a fraction of its computing time: about $100$ hours on NVIDIA-A100 GPUs vs approximately $1.2$ million hours on Intel-KNL CPUs. Notice that the constrained-path approximation violates the variational principle. In contrast, variational Monte Carlo calculations based on the NQS never yield energies below that of the Hamiltonian's ground state. Comparing with the Hartree-Fock approximation, it appears that the hidden-nucleon ansatz captures the overwhelming majority of the correlation energy. 
\begin{figure}[t]
    \includegraphics[width=\columnwidth]{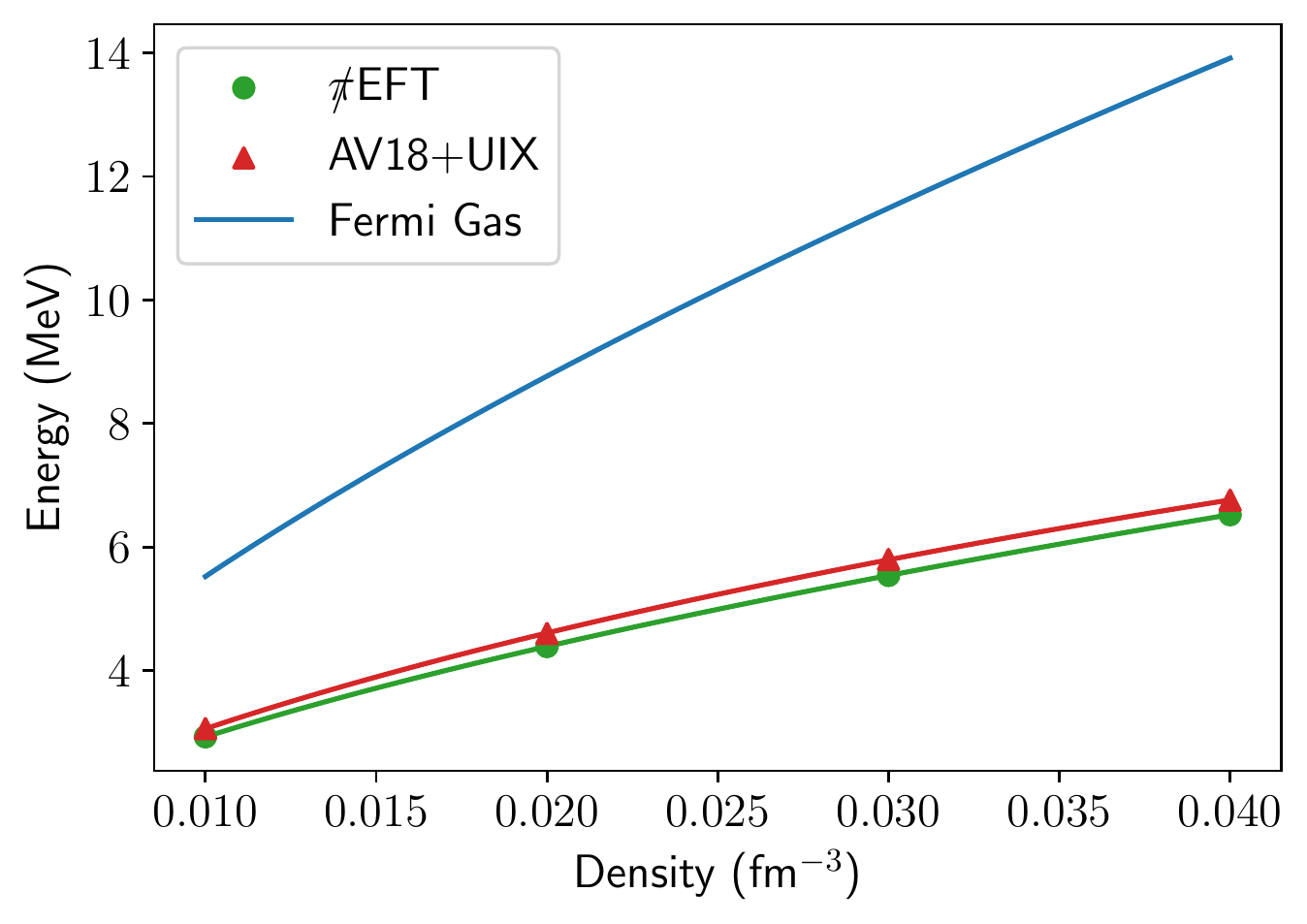}
\caption{Low-density neutron-matter $\slashed{\pi}$EFT equation of state as obtained with the hidden-nucleon NQS  (solid green circles) compared with FHNC/SOC calculations with the AV18+UIX Hamiltonian (red triangles) and the non-interacting Fermi Gas (solid blue line). }
\label{fig:EOS}
\end{figure}

In Fig.~\ref{fig:EOS}, we compare the $\slashed{\pi}$EFT energies obtained with the NQS ansatz against Fermi hypernetted chain/single-operator chain calculations that take as input the sophisticated AV18+UIX Hamiltonian, consistent with the celebrated APR equation of state~\cite{Akmal:1998cf}. For all densities considered, $\slashed{\pi}$EFT and AV18+UIX are in excellent agreement, the maximum difference being always below $0.30$ MeV per particle --- both of them provide energies much below the non-interacting Fermi gas. These minor differences are likely because model ``o'' yields a slightly larger $nn$ scattering length than the experimental value  and, therefore, more attraction in neutron matter. The latter is not compensated for by the three-body force, whose repulsive contribution is at most $0.25$ MeV per neutron. 

\begin{figure}[b]
\includegraphics[width=\columnwidth]{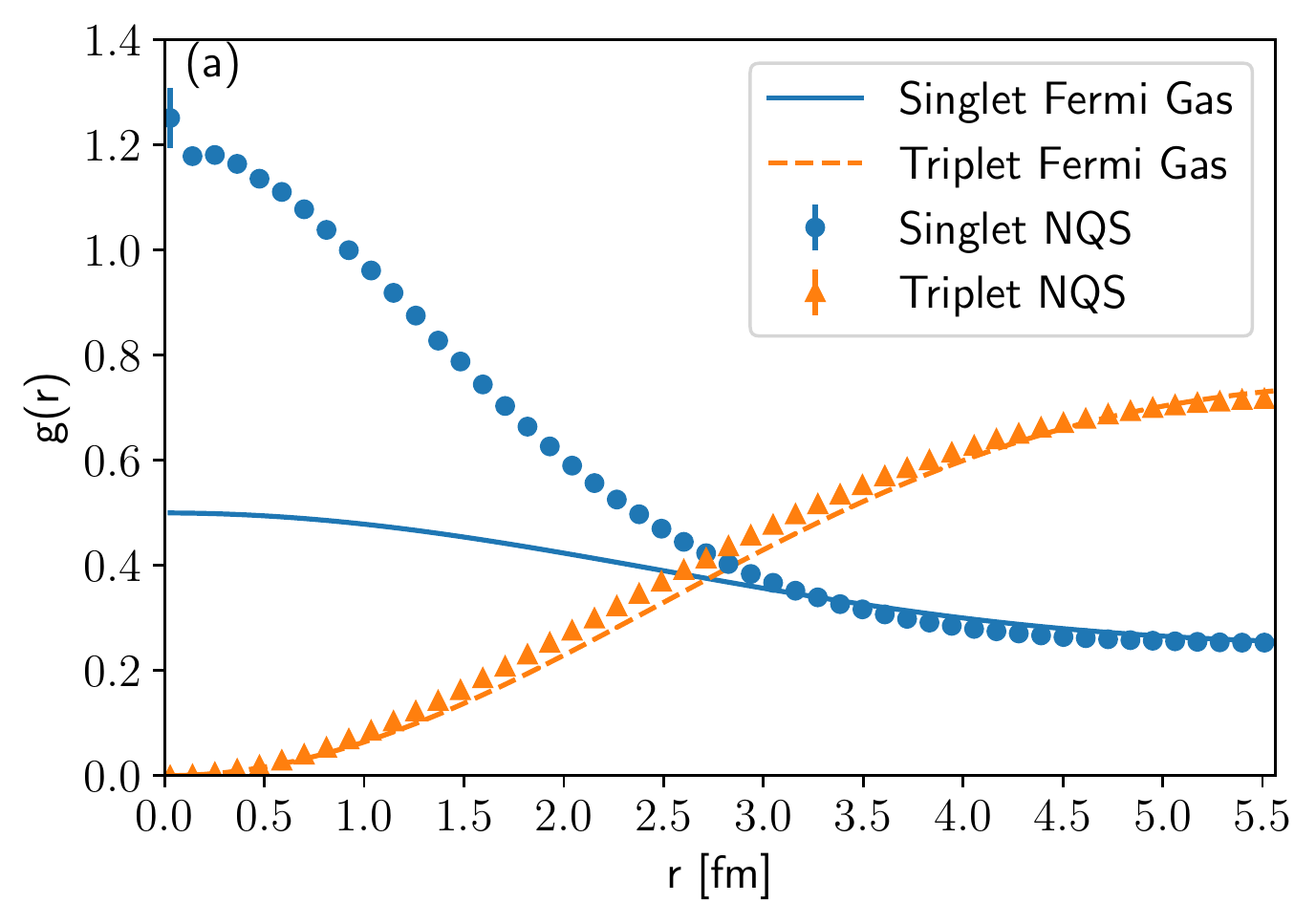}
\label{fig:pair_dist_a}
\includegraphics[width=\columnwidth]{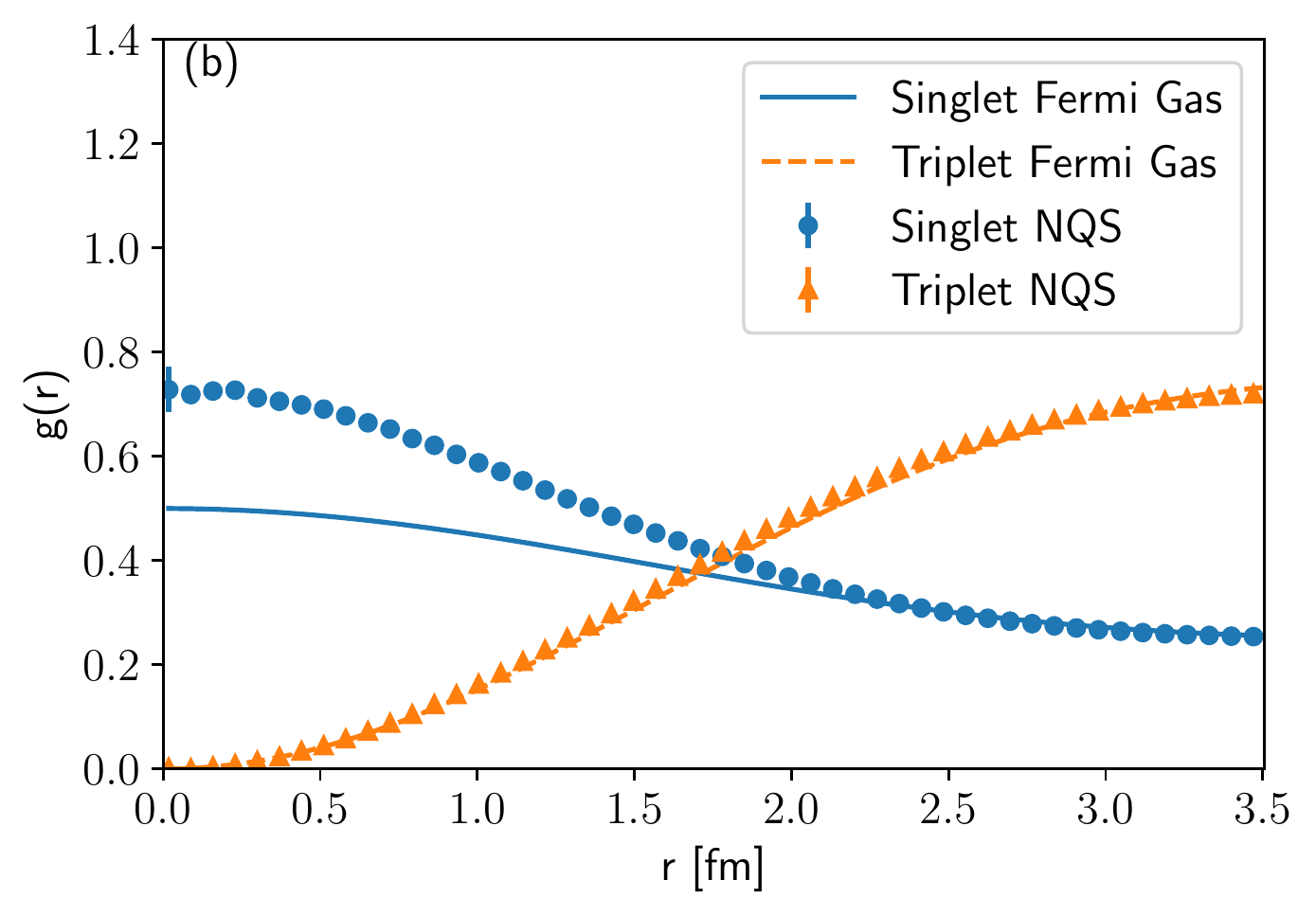}
\label{fig:pair_dist_b}
\caption{Spin-singlet and triplet two-body distribution functions at $\rho=0.01$ fm$^{-3}$ (panel a) and $\rho = 0.04$ fm$^{-3}$ (panel b). The NQS calculations (solid symbols) are compared with non-interacting Fermi Gas results.}
\label{fig:pair_dist}
\end{figure}

Once trained on the systems' energy, the NQS can be used to accurately evaluate a variety of quantum-mechanical observables, such as the spin-singlet and triplet two-body distribution functions defined in Ref.~\cite{Gandolfi2009}. Figure~\ref{fig:pair_dist} shows these distributions at $\rho=0.01$ fm$^{-3}$ (panel a) and $\rho=0.04$ fm$^{-3}$ (panel b). The significant increase in the spin-singlet channel compared to the non-interacting Fermi Gas 
indicates that the NQS wave function can capture the emergence of the $^1S_0$ neutron pairing, despite not being explicitly encoded in the ansatz. Consistent with the behavior of the pairing gap~\cite{Gandolfi:2008id,Benhar:2017mof}, the enhancement is more prominent at $\rho=0.01$ fm$^{-3}$ than $\rho=0.04$ fm$^{-3}$. On the other hand, at these densities, no pairing correlations are present in the spin-triplet channel.

\paragraph{Conclusions --} In this work, we have put forward an NQS suitable to model the normal and superfluid phases of infinite neutron matter in a unified fashion. We improve the expressivity of the hidden-nucleon ansatz of Ref.~\cite{Lovato:2022tjh} by adding state-dependent generalized backflow correlations, whose inclusion has proven beneficial in condensed-matter applications~\cite{Hermann:2019,Pfau:2019}. Periodic-box boundary conditions are imposed by mapping the spatial coordinates of the neutrons onto periodic functions.   

Combined with Monte Carlo techniques to sample the Hilbert space and the stochastic-reconfiguration algorithm to optimize the variational parameters, the NQS yields energies per particle of low-density neutron matter that are in excellent agreement with unconstrained AFDMC calculations at a fraction of the computational cost. In contrast, the computationally-inexpensive AFDMC constrained-path approximation brings about appreciable violations of the variational principle. 

We have shown that $\slashed{\pi}$EFT yields a low-density neutron matter equation of state that is remarkably close to the APR one~\cite{Akmal:1998cf}. This finding paves the way for more systematic comparisons between dilute neutron matter and Fermi gas around the unitary limit. In addition, it enables studies of phenomena relevant to understand the inner crust and the outer core of neutron stars, such as pairing and superfluidity, using relatively simple models of nuclear dynamics.

Finally, we have analyzed the possible onset of Cooper pairing in the neutron medium. Specifically, the NQS two-body distribution functions corresponding to pairs of neutrons in the spin-singlet $^1S_0$ channel exhibit a clear enhancement at small inter-particle distances with respect to the non-interacting case, which is absent in the spin-triplet channel. Consistent with pairing-gap calculations~\cite{Gandolfi:2008id,Benhar:2017mof,Gandolfi:2022dlx}, this behavior is more prominent at smaller densities. Note that this feature has not been encoded in the NQS; rather, it is a self-emerging quantum mechanical phenomenon. 

As a future development, we plan on including more sophisticated interactions, including highly-realistic phenomenological ones such as AV18 + UIX and the local, chiral-EFT potentials of Ref.~\cite{Piarulli:2017dwd,Piarulli:2019pfq,Lovato:2022apd}. The flexibility of the NQS ansatz will also be tested in isospin-asymmetric nucleonic matter at low densities, where strong clustering is expected to occur~\cite{Negele:1971vb}.

\paragraph{Acknowledgments.} We thank R. Wiringa for providing us the AV18+UIX FHNC/SOC energies. We are also grateful to O. Benhar, S. Gandolfi, A. Kievsky, and M. Piarulli for many illuminating discussions. A. L. and B.F. are supported by the U.S. Department of Energy, Office of Science, Office of Nuclear Physics, under contracts DE-AC02-06CH11357, the DOE Early Career Award program, the NUCLEI SciDAC program, and Argonne LDRD awards. J.M.K. and M.H.-J. are supported by the U.S. National Science Foundation Grants No. PHY-1404159 and PHY-2013047. Numerical calculations were performed using resources of the Laboratory Computing Resource Center at Argonne National Laboratory, and the computers of the Argonne Leadership Computing Facility via the ALCC grant ``Short Range Correlations from a Quantum Monte Carlo perspective.''
\bibliography{biblio}

\begin{thebibliography}{50}%
\makeatletter
\providecommand \@ifxundefined [1]{%
 \@ifx{#1\undefined}
}%
\providecommand \@ifnum [1]{%
 \ifnum #1\expandafter \@firstoftwo
 \else \expandafter \@secondoftwo
 \fi
}%
\providecommand \@ifx [1]{%
 \ifx #1\expandafter \@firstoftwo
 \else \expandafter \@secondoftwo
 \fi
}%
\providecommand \natexlab [1]{#1}%
\providecommand \enquote  [1]{``#1''}%
\providecommand \bibnamefont  [1]{#1}%
\providecommand \bibfnamefont [1]{#1}%
\providecommand \citenamefont [1]{#1}%
\providecommand \href@noop [0]{\@secondoftwo}%
\providecommand \href [0]{\begingroup \@sanitize@url \@href}%
\providecommand \@href[1]{\@@startlink{#1}\@@href}%
\providecommand \@@href[1]{\endgroup#1\@@endlink}%
\providecommand \@sanitize@url [0]{\catcode `\\12\catcode `\$12\catcode
  `\&12\catcode `\#12\catcode `\^12\catcode `\_12\catcode `\%12\relax}%
\providecommand \@@startlink[1]{}%
\providecommand \@@endlink[0]{}%
\providecommand \url  [0]{\begingroup\@sanitize@url \@url }%
\providecommand \@url [1]{\endgroup\@href {#1}{\urlprefix }}%
\providecommand \urlprefix  [0]{URL }%
\providecommand \Eprint [0]{\href }%
\providecommand \doibase [0]{http://dx.doi.org/}%
\providecommand \selectlanguage [0]{\@gobble}%
\providecommand \bibinfo  [0]{\@secondoftwo}%
\providecommand \bibfield  [0]{\@secondoftwo}%
\providecommand \translation [1]{[#1]}%
\providecommand \BibitemOpen [0]{}%
\providecommand \bibitemStop [0]{}%
\providecommand \bibitemNoStop [0]{.\EOS\space}%
\providecommand \EOS [0]{\spacefactor3000\relax}%
\providecommand \BibitemShut  [1]{\csname bibitem#1\endcsname}%
\let\auto@bib@innerbib\@empty
\bibitem [{\citenamefont {Abbott}\ \emph
  {et~al.}(2017{\natexlab{a}})\citenamefont {Abbott} \emph
  {et~al.}}]{LIGOScientific:2017vwq}%
  \BibitemOpen
  \bibfield  {author} {\bibinfo {author} {\bibfnamefont {B.~P.}\ \bibnamefont
  {Abbott}} \emph {et~al.} (\bibinfo {collaboration} {LIGO Scientific
  Collaboration and Virgo Collaboration}),\ }\bibfield  {title} {\enquote
  {\bibinfo {title} {{GW170817: Observation of Gravitational Waves from a
  Binary Neutron Star Inspiral}},}\ }\href {\doibase
  10.1103/PhysRevLett.119.161101} {\bibfield  {journal} {\bibinfo  {journal}
  {Phys. Rev. Lett.}\ }\textbf {\bibinfo {volume} {119}},\ \bibinfo {pages}
  {161101} (\bibinfo {year} {2017}{\natexlab{a}})},\ \Eprint
  {http://arxiv.org/abs/1710.05832} {arXiv:1710.05832 [gr-qc]} \BibitemShut
  {NoStop}%
\bibitem [{\citenamefont {Abbott}\ \emph
  {et~al.}(2017{\natexlab{b}})\citenamefont {Abbott} \emph
  {et~al.}}]{Abbott:2017}%
  \BibitemOpen
  \bibfield  {author} {\bibinfo {author} {\bibfnamefont {B.~P.}\ \bibnamefont
  {Abbott}} \emph {et~al.} (\bibinfo {collaboration} {LIGO Scientific
  Collaboration and Virgo Collaboration}),\ }\bibfield  {title} {\enquote
  {\bibinfo {title} {{Multi-messenger Observations of a Binary Neutron Star
  Merger}},}\ }\href {\doibase 10.3847/2041-8213/aa91c9} {\bibfield  {journal}
  {\bibinfo  {journal} {Astrophys. J.}\ }\textbf {\bibinfo {volume} {848}},\
  \bibinfo {pages} {L12} (\bibinfo {year} {2017}{\natexlab{b}})},\ \Eprint
  {http://arxiv.org/abs/1710.05833} {arXiv:1710.05833 [astro-ph.HE]}
  \BibitemShut {NoStop}%
\bibitem [{\citenamefont {Sabatucci}\ and\ \citenamefont
  {Benhar}(2020)}]{Sabatucci:2020xwt}%
  \BibitemOpen
  \bibfield  {author} {\bibinfo {author} {\bibfnamefont {Andrea}\ \bibnamefont
  {Sabatucci}}\ and\ \bibinfo {author} {\bibfnamefont {Omar}\ \bibnamefont
  {Benhar}},\ }\bibfield  {title} {\enquote {\bibinfo {title} {{Tidal
  Deformation of Neutron Stars from Microscopic Models of Nuclear Dynamics}},}\
  }\href {\doibase 10.1103/PhysRevC.101.045807} {\bibfield  {journal} {\bibinfo
   {journal} {Phys. Rev. C}\ }\textbf {\bibinfo {volume} {101}},\ \bibinfo
  {pages} {045807} (\bibinfo {year} {2020})},\ \Eprint
  {http://arxiv.org/abs/2001.06294} {arXiv:2001.06294 [nucl-th]} \BibitemShut
  {NoStop}%
\bibitem [{\citenamefont {Senger}(2021)}]{senger2021}%
  \BibitemOpen
  \bibfield  {author} {\bibinfo {author} {\bibfnamefont {Peter}\ \bibnamefont
  {Senger}},\ }\bibfield  {title} {\enquote {\bibinfo {title} {Probing dense
  nuclear matter in the laboratory: Experiments at fair and nica},}\ }\href
  {\doibase 10.3390/universe7060171} {\bibfield  {journal} {\bibinfo  {journal}
  {Universe}\ }\textbf {\bibinfo {volume} {7}} (\bibinfo {year} {2021}),\
  10.3390/universe7060171}\BibitemShut {NoStop}%
\bibitem [{\citenamefont {Drischler}\ \emph {et~al.}(2016)\citenamefont
  {Drischler}, \citenamefont {Carbone}, \citenamefont {Hebeler},\ and\
  \citenamefont {Schwenk}}]{Drischler:2016djf}%
  \BibitemOpen
  \bibfield  {author} {\bibinfo {author} {\bibfnamefont {C.}~\bibnamefont
  {Drischler}}, \bibinfo {author} {\bibfnamefont {A.}~\bibnamefont {Carbone}},
  \bibinfo {author} {\bibfnamefont {K.}~\bibnamefont {Hebeler}}, \ and\
  \bibinfo {author} {\bibfnamefont {A.}~\bibnamefont {Schwenk}},\ }\bibfield
  {title} {\enquote {\bibinfo {title} {{Neutron matter from chiral two- and
  three-nucleon calculations up to N$^3$LO}},}\ }\href {\doibase
  10.1103/PhysRevC.94.054307} {\bibfield  {journal} {\bibinfo  {journal} {Phys.
  Rev. C}\ }\textbf {\bibinfo {volume} {94}},\ \bibinfo {pages} {054307}
  (\bibinfo {year} {2016})},\ \Eprint {http://arxiv.org/abs/1608.05615}
  {arXiv:1608.05615 [nucl-th]} \BibitemShut {NoStop}%
\bibitem [{\citenamefont {Piarulli}\ \emph {et~al.}(2020)\citenamefont
  {Piarulli}, \citenamefont {Bombaci}, \citenamefont {Logoteta}, \citenamefont
  {Lovato},\ and\ \citenamefont {Wiringa}}]{Piarulli:2019pfq}%
  \BibitemOpen
  \bibfield  {author} {\bibinfo {author} {\bibfnamefont {M.}~\bibnamefont
  {Piarulli}}, \bibinfo {author} {\bibfnamefont {I.}~\bibnamefont {Bombaci}},
  \bibinfo {author} {\bibfnamefont {D.}~\bibnamefont {Logoteta}}, \bibinfo
  {author} {\bibfnamefont {A.}~\bibnamefont {Lovato}}, \ and\ \bibinfo {author}
  {\bibfnamefont {R.~B.}\ \bibnamefont {Wiringa}},\ }\bibfield  {title}
  {\enquote {\bibinfo {title} {{Benchmark calculations of pure neutron matter
  with realistic nucleon-nucleon interactions}},}\ }\href {\doibase
  10.1103/PhysRevC.101.045801} {\bibfield  {journal} {\bibinfo  {journal}
  {Phys. Rev. C}\ }\textbf {\bibinfo {volume} {101}},\ \bibinfo {pages}
  {045801} (\bibinfo {year} {2020})},\ \Eprint
  {http://arxiv.org/abs/1908.04426} {arXiv:1908.04426 [nucl-th]} \BibitemShut
  {NoStop}%
\bibitem [{\citenamefont {Lonardoni}\ \emph {et~al.}(2020)\citenamefont
  {Lonardoni}, \citenamefont {Tews}, \citenamefont {Gandolfi},\ and\
  \citenamefont {Carlson}}]{Lonardoni:2019ypg}%
  \BibitemOpen
  \bibfield  {author} {\bibinfo {author} {\bibfnamefont {D.}~\bibnamefont
  {Lonardoni}}, \bibinfo {author} {\bibfnamefont {I.}~\bibnamefont {Tews}},
  \bibinfo {author} {\bibfnamefont {S.}~\bibnamefont {Gandolfi}}, \ and\
  \bibinfo {author} {\bibfnamefont {J.}~\bibnamefont {Carlson}},\ }\bibfield
  {title} {\enquote {\bibinfo {title} {{Nuclear and neutron-star matter from
  local chiral interactions}},}\ }\href {\doibase
  10.1103/PhysRevResearch.2.022033} {\bibfield  {journal} {\bibinfo  {journal}
  {Phys. Rev. Res.}\ }\textbf {\bibinfo {volume} {2}},\ \bibinfo {pages}
  {022033} (\bibinfo {year} {2020})},\ \Eprint
  {http://arxiv.org/abs/1912.09411} {arXiv:1912.09411 [nucl-th]} \BibitemShut
  {NoStop}%
\bibitem [{\citenamefont {Jiang}\ \emph {et~al.}(2020)\citenamefont {Jiang},
  \citenamefont {Ekstr\"om}, \citenamefont {Forss\'en}, \citenamefont {Hagen},
  \citenamefont {Jansen},\ and\ \citenamefont {Papenbrock}}]{Jiang:2020the}%
  \BibitemOpen
  \bibfield  {author} {\bibinfo {author} {\bibfnamefont {W.~G.}\ \bibnamefont
  {Jiang}}, \bibinfo {author} {\bibfnamefont {A.}~\bibnamefont {Ekstr\"om}},
  \bibinfo {author} {\bibfnamefont {C.}~\bibnamefont {Forss\'en}}, \bibinfo
  {author} {\bibfnamefont {G.}~\bibnamefont {Hagen}}, \bibinfo {author}
  {\bibfnamefont {G.~R.}\ \bibnamefont {Jansen}}, \ and\ \bibinfo {author}
  {\bibfnamefont {T.}~\bibnamefont {Papenbrock}},\ }\bibfield  {title}
  {\enquote {\bibinfo {title} {{Accurate bulk properties of nuclei from $A=2$
  to $\infty$ from potentials with $\Delta$ isobars}},}\ }\href {\doibase
  10.1103/PhysRevC.102.054301} {\bibfield  {journal} {\bibinfo  {journal}
  {Phys. Rev. C}\ }\textbf {\bibinfo {volume} {102}},\ \bibinfo {pages}
  {054301} (\bibinfo {year} {2020})},\ \Eprint
  {http://arxiv.org/abs/2006.16774} {arXiv:2006.16774 [nucl-th]} \BibitemShut
  {NoStop}%
\bibitem [{\citenamefont {Sammarruca}\ and\ \citenamefont
  {Millerson}(2021)}]{Sammarruca:2021bpn}%
  \BibitemOpen
  \bibfield  {author} {\bibinfo {author} {\bibfnamefont {Francesca}\
  \bibnamefont {Sammarruca}}\ and\ \bibinfo {author} {\bibfnamefont {Randy}\
  \bibnamefont {Millerson}},\ }\bibfield  {title} {\enquote {\bibinfo {title}
  {{Overview of symmetric nuclear matter properties from chiral interactions up
  to fourth order of the chiral expansion}},}\ }\href {\doibase
  10.1103/PhysRevC.104.064312} {\bibfield  {journal} {\bibinfo  {journal}
  {Phys. Rev. C}\ }\textbf {\bibinfo {volume} {104}},\ \bibinfo {pages}
  {064312} (\bibinfo {year} {2021})},\ \Eprint
  {http://arxiv.org/abs/2109.01985} {arXiv:2109.01985 [nucl-th]} \BibitemShut
  {NoStop}%
\bibitem [{\citenamefont {Heiselberg}\ and\ \citenamefont
  {Hjorth-Jensen}(2000)}]{Heiselberg2000}%
  \BibitemOpen
  \bibfield  {author} {\bibinfo {author} {\bibfnamefont {H.}~\bibnamefont
  {Heiselberg}}\ and\ \bibinfo {author} {\bibfnamefont {M.}~\bibnamefont
  {Hjorth-Jensen}},\ }\bibfield  {title} {\enquote {\bibinfo {title} {Phases of
  dense matter in neutron stars},}\ }\href {\doibase
  https://doi.org/10.1016/S0370-1573(99)00110-6} {\bibfield  {journal}
  {\bibinfo  {journal} {Phys. Rep.}\ }\textbf {\bibinfo {volume} {328}},\
  \bibinfo {pages} {237} (\bibinfo {year} {2000})}\BibitemShut {NoStop}%
\bibitem [{\citenamefont {Sabatucci}\ \emph {et~al.}(2022)\citenamefont
  {Sabatucci}, \citenamefont {Benhar}, \citenamefont {Maselli},\ and\
  \citenamefont {Pacilio}}]{Sabatucci:2022qyi}%
  \BibitemOpen
  \bibfield  {author} {\bibinfo {author} {\bibfnamefont {Andrea}\ \bibnamefont
  {Sabatucci}}, \bibinfo {author} {\bibfnamefont {Omar}\ \bibnamefont
  {Benhar}}, \bibinfo {author} {\bibfnamefont {Andrea}\ \bibnamefont
  {Maselli}}, \ and\ \bibinfo {author} {\bibfnamefont {Costantino}\
  \bibnamefont {Pacilio}},\ }\bibfield  {title} {\enquote {\bibinfo {title}
  {{Sensitivity of neutron star observations to three-nucleon forces}},}\
  }\href {\doibase 10.1103/PhysRevD.106.083010} {\bibfield  {journal} {\bibinfo
   {journal} {Phys. Rev. D}\ }\textbf {\bibinfo {volume} {106}},\ \bibinfo
  {pages} {083010} (\bibinfo {year} {2022})},\ \Eprint
  {http://arxiv.org/abs/2206.11286} {arXiv:2206.11286 [astro-ph.HE]}
  \BibitemShut {NoStop}%
\bibitem [{\citenamefont {Sedrakian}\ \emph {et~al.}(2006)\citenamefont
  {Sedrakian}, \citenamefont {Clark},\ and\ \citenamefont
  {Alford}}]{Sedrakian:2006}%
  \BibitemOpen
  \bibfield  {author} {\bibinfo {author} {\bibfnamefont {Armen}\ \bibnamefont
  {Sedrakian}}, \bibinfo {author} {\bibfnamefont {John~W}\ \bibnamefont
  {Clark}}, \ and\ \bibinfo {author} {\bibfnamefont {Mark}\ \bibnamefont
  {Alford}},\ }\href {\doibase 10.1142/6194} {\emph {\bibinfo {title} {Pairing
  in Fermionic Systems}}}\ (\bibinfo  {publisher} {WORLD SCIENTIFIC},\ \bibinfo
  {year} {2006})\ \Eprint
  {http://arxiv.org/abs/https://www.worldscientific.com/doi/pdf/10.1142/6194}
  {https://www.worldscientific.com/doi/pdf/10.1142/6194} \BibitemShut {NoStop}%
\bibitem [{\citenamefont {Benhar}\ and\ \citenamefont
  {De~Rosi}(2017)}]{Benhar:2017mof}%
  \BibitemOpen
  \bibfield  {author} {\bibinfo {author} {\bibfnamefont {Omar}\ \bibnamefont
  {Benhar}}\ and\ \bibinfo {author} {\bibfnamefont {Giulia}\ \bibnamefont
  {De~Rosi}},\ }\bibfield  {title} {\enquote {\bibinfo {title} {{Superfluid Gap
  in Neutron Matter from a Microscopic Effective Interaction}},}\ }\href
  {\doibase 10.1007/s10909-017-1823-x} {\bibfield  {journal} {\bibinfo
  {journal} {J. Low Temp. Phys.}\ }\textbf {\bibinfo {volume} {189}},\ \bibinfo
  {pages} {250--261} (\bibinfo {year} {2017})},\ \Eprint
  {http://arxiv.org/abs/1705.06607} {arXiv:1705.06607 [nucl-th]} \BibitemShut
  {NoStop}%
\bibitem [{\citenamefont {Dean}\ and\ \citenamefont
  {Hjorth-Jensen}(2003)}]{Dean2003}%
  \BibitemOpen
  \bibfield  {author} {\bibinfo {author} {\bibfnamefont {D.~J.}\ \bibnamefont
  {Dean}}\ and\ \bibinfo {author} {\bibfnamefont {M.}~\bibnamefont
  {Hjorth-Jensen}},\ }\bibfield  {title} {\enquote {\bibinfo {title} {Pairing
  in nuclear systems: from neutron stars to finite nuclei},}\ }\href {\doibase
  10.1103/RevModPhys.75.607} {\bibfield  {journal} {\bibinfo  {journal} {Rev.
  Mod. Phys.}\ }\textbf {\bibinfo {volume} {75}},\ \bibinfo {pages} {607}
  (\bibinfo {year} {2003})}\BibitemShut {NoStop}%
\bibitem [{\citenamefont {Yakovlev}\ and\ \citenamefont
  {Pethick}(2004)}]{Yakovlev:2004iq}%
  \BibitemOpen
  \bibfield  {author} {\bibinfo {author} {\bibfnamefont {Dima~G.}\ \bibnamefont
  {Yakovlev}}\ and\ \bibinfo {author} {\bibfnamefont {C.~J.}\ \bibnamefont
  {Pethick}},\ }\bibfield  {title} {\enquote {\bibinfo {title} {{Neutron star
  cooling}},}\ }\href {\doibase 10.1146/annurev.astro.42.053102.134013}
  {\bibfield  {journal} {\bibinfo  {journal} {Ann. Rev. Astron. Astrophys.}\
  }\textbf {\bibinfo {volume} {42}},\ \bibinfo {pages} {169--210} (\bibinfo
  {year} {2004})},\ \Eprint {http://arxiv.org/abs/astro-ph/0402143}
  {arXiv:astro-ph/0402143} \BibitemShut {NoStop}%
\bibitem [{\citenamefont {Page}\ \emph {et~al.}(2011)\citenamefont {Page},
  \citenamefont {Prakash}, \citenamefont {Lattimer},\ and\ \citenamefont
  {Steiner}}]{Page:2010aw}%
  \BibitemOpen
  \bibfield  {author} {\bibinfo {author} {\bibfnamefont {Dany}\ \bibnamefont
  {Page}}, \bibinfo {author} {\bibfnamefont {Madappa}\ \bibnamefont {Prakash}},
  \bibinfo {author} {\bibfnamefont {James~M.}\ \bibnamefont {Lattimer}}, \ and\
  \bibinfo {author} {\bibfnamefont {Andrew~W.}\ \bibnamefont {Steiner}},\
  }\bibfield  {title} {\enquote {\bibinfo {title} {{Rapid Cooling of the
  Neutron Star in Cassiopeia A Triggered by Neutron Superfluidity in Dense
  Matter}},}\ }\href {\doibase 10.1103/PhysRevLett.106.081101} {\bibfield
  {journal} {\bibinfo  {journal} {Phys. Rev. Lett.}\ }\textbf {\bibinfo
  {volume} {106}},\ \bibinfo {pages} {081101} (\bibinfo {year} {2011})},\
  \Eprint {http://arxiv.org/abs/1011.6142} {arXiv:1011.6142 [astro-ph.HE]}
  \BibitemShut {NoStop}%
\bibitem [{\citenamefont {Monrozeau}\ \emph {et~al.}(2007)\citenamefont
  {Monrozeau}, \citenamefont {Margueron},\ and\ \citenamefont
  {Sandulescu}}]{Monrozeau:2007xu}%
  \BibitemOpen
  \bibfield  {author} {\bibinfo {author} {\bibfnamefont {C.}~\bibnamefont
  {Monrozeau}}, \bibinfo {author} {\bibfnamefont {J.}~\bibnamefont
  {Margueron}}, \ and\ \bibinfo {author} {\bibfnamefont {N.}~\bibnamefont
  {Sandulescu}},\ }\bibfield  {title} {\enquote {\bibinfo {title} {{Nuclear
  superfluidity and cooling time of neutron-star crust}},}\ }\href {\doibase
  10.1103/PhysRevC.75.065807} {\bibfield  {journal} {\bibinfo  {journal} {Phys.
  Rev. C}\ }\textbf {\bibinfo {volume} {75}},\ \bibinfo {pages} {065807}
  (\bibinfo {year} {2007})},\ \Eprint {http://arxiv.org/abs/nucl-th/0703064}
  {arXiv:nucl-th/0703064} \BibitemShut {NoStop}%
\bibitem [{\citenamefont {Nowacki}\ \emph {et~al.}(2021)\citenamefont
  {Nowacki}, \citenamefont {Obertelli},\ and\ \citenamefont
  {Poves}}]{Nowacki:2021fjw}%
  \BibitemOpen
  \bibfield  {author} {\bibinfo {author} {\bibfnamefont {Fr\'ed\'eric}\
  \bibnamefont {Nowacki}}, \bibinfo {author} {\bibfnamefont {Alexandre}\
  \bibnamefont {Obertelli}}, \ and\ \bibinfo {author} {\bibfnamefont {Alfredo}\
  \bibnamefont {Poves}},\ }\bibfield  {title} {\enquote {\bibinfo {title} {{The
  neutron-rich edge of the nuclear landscape: Experiment and theory.}}}\ }\href
  {\doibase 10.1016/j.ppnp.2021.103866} {\bibfield  {journal} {\bibinfo
  {journal} {Prog. Part. Nucl. Phys.}\ }\textbf {\bibinfo {volume} {120}},\
  \bibinfo {pages} {103866} (\bibinfo {year} {2021})},\ \Eprint
  {http://arxiv.org/abs/2104.06238} {arXiv:2104.06238 [nucl-th]} \BibitemShut
  {NoStop}%
\bibitem [{\citenamefont {Carlson}\ \emph {et~al.}(2015)\citenamefont
  {Carlson}, \citenamefont {Gandolfi}, \citenamefont {Pederiva}, \citenamefont
  {Pieper}, \citenamefont {Schiavilla}, \citenamefont {Schmidt},\ and\
  \citenamefont {Wiringa}}]{Carlson:2014vla}%
  \BibitemOpen
  \bibfield  {author} {\bibinfo {author} {\bibfnamefont {J.}~\bibnamefont
  {Carlson}}, \bibinfo {author} {\bibfnamefont {S.}~\bibnamefont {Gandolfi}},
  \bibinfo {author} {\bibfnamefont {F.}~\bibnamefont {Pederiva}}, \bibinfo
  {author} {\bibfnamefont {Steven~C.}\ \bibnamefont {Pieper}}, \bibinfo
  {author} {\bibfnamefont {R.}~\bibnamefont {Schiavilla}}, \bibinfo {author}
  {\bibfnamefont {K.~E.}\ \bibnamefont {Schmidt}}, \ and\ \bibinfo {author}
  {\bibfnamefont {R.~B.}\ \bibnamefont {Wiringa}},\ }\bibfield  {title}
  {\enquote {\bibinfo {title} {{Quantum Monte Carlo methods for nuclear
  physics}},}\ }\href {\doibase 10.1103/RevModPhys.87.1067} {\bibfield
  {journal} {\bibinfo  {journal} {Rev. Mod. Phys.}\ }\textbf {\bibinfo {volume}
  {87}},\ \bibinfo {pages} {1067} (\bibinfo {year} {2015})},\ \Eprint
  {http://arxiv.org/abs/1412.3081} {arXiv:1412.3081 [nucl-th]} \BibitemShut
  {NoStop}%
\bibitem [{\citenamefont {Schmidt}\ and\ \citenamefont
  {Fantoni}(1999)}]{Schmidt:1999lik}%
  \BibitemOpen
  \bibfield  {author} {\bibinfo {author} {\bibfnamefont {K.~E.}\ \bibnamefont
  {Schmidt}}\ and\ \bibinfo {author} {\bibfnamefont {S.}~\bibnamefont
  {Fantoni}},\ }\bibfield  {title} {\enquote {\bibinfo {title} {{A quantum
  Monte Carlo method for nucleon systems}},}\ }\href {\doibase
  10.1016/S0370-2693(98)01522-6} {\bibfield  {journal} {\bibinfo  {journal}
  {Phys. Lett. B}\ }\textbf {\bibinfo {volume} {446}},\ \bibinfo {pages}
  {99--103} (\bibinfo {year} {1999})}\BibitemShut {NoStop}%
\bibitem [{\citenamefont {Lovato}\ \emph
  {et~al.}(2022{\natexlab{a}})\citenamefont {Lovato}, \citenamefont {Bombaci},
  \citenamefont {Logoteta}, \citenamefont {Piarulli},\ and\ \citenamefont
  {Wiringa}}]{Lovato:2022apd}%
  \BibitemOpen
  \bibfield  {author} {\bibinfo {author} {\bibfnamefont {A.}~\bibnamefont
  {Lovato}}, \bibinfo {author} {\bibfnamefont {I.}~\bibnamefont {Bombaci}},
  \bibinfo {author} {\bibfnamefont {D.}~\bibnamefont {Logoteta}}, \bibinfo
  {author} {\bibfnamefont {M.}~\bibnamefont {Piarulli}}, \ and\ \bibinfo
  {author} {\bibfnamefont {R.~B.}\ \bibnamefont {Wiringa}},\ }\bibfield
  {title} {\enquote {\bibinfo {title} {{Benchmark calculations of infinite
  neutron matter with realistic two- and three-nucleon potentials}},}\ }\href
  {\doibase 10.1103/PhysRevC.105.055808} {\bibfield  {journal} {\bibinfo
  {journal} {Phys. Rev. C}\ }\textbf {\bibinfo {volume} {105}},\ \bibinfo
  {pages} {055808} (\bibinfo {year} {2022}{\natexlab{a}})},\ \Eprint
  {http://arxiv.org/abs/2202.10293} {arXiv:2202.10293 [nucl-th]} \BibitemShut
  {NoStop}%
\bibitem [{\citenamefont {Gandolfi}\ \emph {et~al.}(2008)\citenamefont
  {Gandolfi}, \citenamefont {Illarionov}, \citenamefont {Fantoni},
  \citenamefont {Pederiva},\ and\ \citenamefont {Schmidt}}]{Gandolfi:2008id}%
  \BibitemOpen
  \bibfield  {author} {\bibinfo {author} {\bibfnamefont {S.}~\bibnamefont
  {Gandolfi}}, \bibinfo {author} {\bibfnamefont {A.~Yu.}\ \bibnamefont
  {Illarionov}}, \bibinfo {author} {\bibfnamefont {S.}~\bibnamefont {Fantoni}},
  \bibinfo {author} {\bibfnamefont {F.}~\bibnamefont {Pederiva}}, \ and\
  \bibinfo {author} {\bibfnamefont {K.~E.}\ \bibnamefont {Schmidt}},\
  }\bibfield  {title} {\enquote {\bibinfo {title} {{Equation of state of
  superfluid neutron matter and the calculation of S(0)-1 pairing gap}},}\
  }\href {\doibase 10.1103/PhysRevLett.101.132501} {\bibfield  {journal}
  {\bibinfo  {journal} {Phys. Rev. Lett.}\ }\textbf {\bibinfo {volume} {101}},\
  \bibinfo {pages} {132501} (\bibinfo {year} {2008})},\ \Eprint
  {http://arxiv.org/abs/0805.2513} {arXiv:0805.2513 [nucl-th]} \BibitemShut
  {NoStop}%
\bibitem [{\citenamefont {Gandolfi}\ \emph {et~al.}(2022)\citenamefont
  {Gandolfi}, \citenamefont {Palkanoglou}, \citenamefont {Carlson},
  \citenamefont {Gezerlis},\ and\ \citenamefont {Schmidt}}]{Gandolfi:2022dlx}%
  \BibitemOpen
  \bibfield  {author} {\bibinfo {author} {\bibfnamefont {Stefano}\ \bibnamefont
  {Gandolfi}}, \bibinfo {author} {\bibfnamefont {Georgios}\ \bibnamefont
  {Palkanoglou}}, \bibinfo {author} {\bibfnamefont {Joseph}\ \bibnamefont
  {Carlson}}, \bibinfo {author} {\bibfnamefont {Alexandros}\ \bibnamefont
  {Gezerlis}}, \ and\ \bibinfo {author} {\bibfnamefont {Kevin~E.}\ \bibnamefont
  {Schmidt}},\ }\bibfield  {title} {\enquote {\bibinfo {title} {{The 1S0
  Pairing Gap in Neutron Matter}},}\ }\href {\doibase 10.3390/condmat7010019}
  {\bibfield  {journal} {\bibinfo  {journal} {Condens. Mat.}\ }\textbf
  {\bibinfo {volume} {7}},\ \bibinfo {pages} {19} (\bibinfo {year} {2022})},\
  \Eprint {http://arxiv.org/abs/2201.01308} {arXiv:2201.01308 [nucl-th]}
  \BibitemShut {NoStop}%
\bibitem [{\citenamefont {Bajdich}\ \emph {et~al.}(2006)\citenamefont
  {Bajdich}, \citenamefont {Mitas}, \citenamefont {Drobny}, \citenamefont
  {Wagner},\ and\ \citenamefont {Schmidt}}]{Bajdich:2006zz}%
  \BibitemOpen
  \bibfield  {author} {\bibinfo {author} {\bibfnamefont {M.}~\bibnamefont
  {Bajdich}}, \bibinfo {author} {\bibfnamefont {L.}~\bibnamefont {Mitas}},
  \bibinfo {author} {\bibfnamefont {G.}~\bibnamefont {Drobny}}, \bibinfo
  {author} {\bibfnamefont {L.~K.}\ \bibnamefont {Wagner}}, \ and\ \bibinfo
  {author} {\bibfnamefont {K.~E.}\ \bibnamefont {Schmidt}},\ }\bibfield
  {title} {\enquote {\bibinfo {title} {{Pfaffian pairing wave functions in
  electronic structure quantum Monte Carlo}},}\ }\href {\doibase
  10.1103/PhysRevLett.96.130201} {\bibfield  {journal} {\bibinfo  {journal}
  {Phys. Rev. Lett.}\ }\textbf {\bibinfo {volume} {96}},\ \bibinfo {pages}
  {130201} (\bibinfo {year} {2006})},\ \Eprint
  {http://arxiv.org/abs/cond-mat/0512327} {arXiv:cond-mat/0512327} \BibitemShut
  {NoStop}%
\bibitem [{\citenamefont {Carleo}\ and\ \citenamefont
  {Troyer}(2017)}]{carleo_solving_2017}%
  \BibitemOpen
  \bibfield  {author} {\bibinfo {author} {\bibfnamefont {Giuseppe}\
  \bibnamefont {Carleo}}\ and\ \bibinfo {author} {\bibfnamefont {Matthias}\
  \bibnamefont {Troyer}},\ }\bibfield  {title} {\enquote {\bibinfo {title}
  {Solving the quantum many-body problem with artificial neural networks},}\
  }\href {\doibase 10.1126/science.aag2302} {\bibfield  {journal} {\bibinfo
  {journal} {Science}\ }\textbf {\bibinfo {volume} {355}},\ \bibinfo {pages}
  {602--606} (\bibinfo {year} {2017})}\BibitemShut {NoStop}%
\bibitem [{\citenamefont {Keeble}\ and\ \citenamefont
  {Rios}(2020)}]{Keeble:2019bkv}%
  \BibitemOpen
  \bibfield  {author} {\bibinfo {author} {\bibfnamefont {J.~W.~T.}\
  \bibnamefont {Keeble}}\ and\ \bibinfo {author} {\bibfnamefont
  {A.}~\bibnamefont {Rios}},\ }\bibfield  {title} {\enquote {\bibinfo {title}
  {{Machine learning the deuteron}},}\ }\href {\doibase
  10.1016/j.physletb.2020.135743} {\bibfield  {journal} {\bibinfo  {journal}
  {Phys. Lett. B}\ }\textbf {\bibinfo {volume} {809}},\ \bibinfo {pages}
  {135743} (\bibinfo {year} {2020})},\ \Eprint
  {http://arxiv.org/abs/1911.13092} {arXiv:1911.13092 [nucl-th]} \BibitemShut
  {NoStop}%
\bibitem [{\citenamefont {Adams}\ \emph {et~al.}(2021)\citenamefont {Adams},
  \citenamefont {Carleo}, \citenamefont {Lovato},\ and\ \citenamefont
  {Rocco}}]{Adams:2020aax}%
  \BibitemOpen
  \bibfield  {author} {\bibinfo {author} {\bibfnamefont {Corey}\ \bibnamefont
  {Adams}}, \bibinfo {author} {\bibfnamefont {Giuseppe}\ \bibnamefont
  {Carleo}}, \bibinfo {author} {\bibfnamefont {Alessandro}\ \bibnamefont
  {Lovato}}, \ and\ \bibinfo {author} {\bibfnamefont {Noemi}\ \bibnamefont
  {Rocco}},\ }\bibfield  {title} {\enquote {\bibinfo {title} {{Variational
  Monte Carlo Calculations of A\ensuremath{\leq}4 Nuclei with an Artificial
  Neural-Network Correlator Ansatz}},}\ }\href {\doibase
  10.1103/PhysRevLett.127.022502} {\bibfield  {journal} {\bibinfo  {journal}
  {Phys. Rev. Lett.}\ }\textbf {\bibinfo {volume} {127}},\ \bibinfo {pages}
  {022502} (\bibinfo {year} {2021})},\ \Eprint
  {http://arxiv.org/abs/2007.14282} {arXiv:2007.14282 [nucl-th]} \BibitemShut
  {NoStop}%
\bibitem [{\citenamefont {Gnech}\ \emph {et~al.}(2022)\citenamefont {Gnech},
  \citenamefont {Adams}, \citenamefont {Brawand}, \citenamefont {Carleo},
  \citenamefont {Lovato},\ and\ \citenamefont {Rocco}}]{Gnech:2021wfn}%
  \BibitemOpen
  \bibfield  {author} {\bibinfo {author} {\bibfnamefont {Alex}\ \bibnamefont
  {Gnech}}, \bibinfo {author} {\bibfnamefont {Corey}\ \bibnamefont {Adams}},
  \bibinfo {author} {\bibfnamefont {Nicholas}\ \bibnamefont {Brawand}},
  \bibinfo {author} {\bibfnamefont {Giuseppe}\ \bibnamefont {Carleo}}, \bibinfo
  {author} {\bibfnamefont {Alessandro}\ \bibnamefont {Lovato}}, \ and\ \bibinfo
  {author} {\bibfnamefont {Noemi}\ \bibnamefont {Rocco}},\ }\bibfield  {title}
  {\enquote {\bibinfo {title} {{Nuclei with up to $\boldsymbol{A=6}$ nucleons
  with artificial neural network wave functions}},}\ }\href {\doibase
  10.1007/s00601-021-01706-0} {\bibfield  {journal} {\bibinfo  {journal} {Few
  Body Syst.}\ }\textbf {\bibinfo {volume} {63}},\ \bibinfo {pages} {7}
  (\bibinfo {year} {2022})},\ \Eprint {http://arxiv.org/abs/2108.06836}
  {arXiv:2108.06836 [nucl-th]} \BibitemShut {NoStop}%
\bibitem [{\citenamefont {Lovato}\ \emph
  {et~al.}(2022{\natexlab{b}})\citenamefont {Lovato}, \citenamefont {Adams},
  \citenamefont {Carleo},\ and\ \citenamefont {Rocco}}]{Lovato:2022tjh}%
  \BibitemOpen
  \bibfield  {author} {\bibinfo {author} {\bibfnamefont {A.}~\bibnamefont
  {Lovato}}, \bibinfo {author} {\bibfnamefont {C.}~\bibnamefont {Adams}},
  \bibinfo {author} {\bibfnamefont {G.}~\bibnamefont {Carleo}}, \ and\ \bibinfo
  {author} {\bibfnamefont {N.}~\bibnamefont {Rocco}},\ }\bibfield  {title}
  {\enquote {\bibinfo {title} {{Hidden-nucleons neural-network quantum states
  for the nuclear many-body problem}},}\ }\href@noop {} {\  (\bibinfo {year}
  {2022}{\natexlab{b}})},\ \Eprint {http://arxiv.org/abs/2206.10021}
  {arXiv:2206.10021 [nucl-th]} \BibitemShut {NoStop}%
\bibitem [{\citenamefont {Yang}\ and\ \citenamefont
  {Zhao}(2022)}]{Yang:2022esu}%
  \BibitemOpen
  \bibfield  {author} {\bibinfo {author} {\bibfnamefont {Y.~L.}\ \bibnamefont
  {Yang}}\ and\ \bibinfo {author} {\bibfnamefont {P.~W.}\ \bibnamefont
  {Zhao}},\ }\bibfield  {title} {\enquote {\bibinfo {title} {{A consistent
  description of the relativistic effects and three-body interactions in atomic
  nuclei}},}\ }\href {\doibase 10.1016/j.physletb.2022.137587} {\bibfield
  {journal} {\bibinfo  {journal} {Phys. Lett. B}\ }\textbf {\bibinfo {volume}
  {835}},\ \bibinfo {pages} {137587} (\bibinfo {year} {2022})},\ \Eprint
  {http://arxiv.org/abs/2206.13208} {arXiv:2206.13208 [nucl-th]} \BibitemShut
  {NoStop}%
\bibitem [{\citenamefont {Rigo}\ \emph {et~al.}(2022)\citenamefont {Rigo},
  \citenamefont {Hall}, \citenamefont {Hjorth-Jensen}, \citenamefont {Lovato},\
  and\ \citenamefont {Pederiva}}]{Rigo:2022ces}%
  \BibitemOpen
  \bibfield  {author} {\bibinfo {author} {\bibfnamefont {Mauro}\ \bibnamefont
  {Rigo}}, \bibinfo {author} {\bibfnamefont {Benjamin}\ \bibnamefont {Hall}},
  \bibinfo {author} {\bibfnamefont {Morten}\ \bibnamefont {Hjorth-Jensen}},
  \bibinfo {author} {\bibfnamefont {Alessandro}\ \bibnamefont {Lovato}}, \ and\
  \bibinfo {author} {\bibfnamefont {Francesco}\ \bibnamefont {Pederiva}},\
  }\bibfield  {title} {\enquote {\bibinfo {title} {{Solving the nuclear pairing
  model with neural network quantum states}},}\ }\href@noop {} {\  (\bibinfo
  {year} {2022})},\ \Eprint {http://arxiv.org/abs/2211.04614} {arXiv:2211.04614
  [nucl-th]} \BibitemShut {NoStop}%
\bibitem [{\citenamefont {{Hermann}}\ \emph {et~al.}(2020)\citenamefont
  {{Hermann}}, \citenamefont {{Sch{\"a}tzle}},\ and\ \citenamefont
  {{No{\'e}}}}]{Hermann:2019}%
  \BibitemOpen
  \bibfield  {author} {\bibinfo {author} {\bibfnamefont {Jan}\ \bibnamefont
  {{Hermann}}}, \bibinfo {author} {\bibfnamefont {Zeno}\ \bibnamefont
  {{Sch{\"a}tzle}}}, \ and\ \bibinfo {author} {\bibfnamefont {Frank}\
  \bibnamefont {{No{\'e}}}},\ }\bibfield  {title} {\enquote {\bibinfo {title}
  {{Deep-neural-network solution of the electronic Schr{\"o}dinger
  equation}},}\ }\href {\doibase 10.1038/s41557-020-0544-y} {\bibfield
  {journal} {\bibinfo  {journal} {Nature Chemistry}\ }\textbf {\bibinfo
  {volume} {12}},\ \bibinfo {pages} {891--897} (\bibinfo {year}
  {2020})}\BibitemShut {NoStop}%
\bibitem [{\citenamefont {{Pfau}}\ \emph {et~al.}(2020)\citenamefont {{Pfau}},
  \citenamefont {{Spencer}}, \citenamefont {{Matthews}},\ and\ \citenamefont
  {{Foulkes}}}]{Pfau:2019}%
  \BibitemOpen
  \bibfield  {author} {\bibinfo {author} {\bibfnamefont {David}\ \bibnamefont
  {{Pfau}}}, \bibinfo {author} {\bibfnamefont {James~S.}\ \bibnamefont
  {{Spencer}}}, \bibinfo {author} {\bibfnamefont {Alexander G.~D.~G.}\
  \bibnamefont {{Matthews}}}, \ and\ \bibinfo {author} {\bibfnamefont
  {W.~M.~C.}\ \bibnamefont {{Foulkes}}},\ }\bibfield  {title} {\enquote
  {\bibinfo {title} {{Ab initio solution of the many-electron Schr{\"o}dinger
  equation with deep neural networks}},}\ }\href {\doibase
  10.1103/PhysRevResearch.2.033429} {\bibfield  {journal} {\bibinfo  {journal}
  {Physical Review Research}\ }\textbf {\bibinfo {volume} {2}},\ \bibinfo {eid}
  {033429} (\bibinfo {year} {2020})},\ \Eprint
  {http://arxiv.org/abs/1909.02487} {arXiv:1909.02487 [physics.chem-ph]}
  \BibitemShut {NoStop}%
\bibitem [{\citenamefont {Brualla}\ \emph {et~al.}(2003)\citenamefont
  {Brualla}, \citenamefont {Fantoni}, \citenamefont {Sarsa}, \citenamefont
  {Schmidt},\ and\ \citenamefont {Vitiello}}]{Brualla:2003gw}%
  \BibitemOpen
  \bibfield  {author} {\bibinfo {author} {\bibfnamefont {L.}~\bibnamefont
  {Brualla}}, \bibinfo {author} {\bibfnamefont {S.}~\bibnamefont {Fantoni}},
  \bibinfo {author} {\bibfnamefont {A.}~\bibnamefont {Sarsa}}, \bibinfo
  {author} {\bibfnamefont {K.~E}\ \bibnamefont {Schmidt}}, \ and\ \bibinfo
  {author} {\bibfnamefont {S.~A.}\ \bibnamefont {Vitiello}},\ }\bibfield
  {title} {\enquote {\bibinfo {title} {{Spin orbit induced backflow in neutron
  matter with auxiliary field diffusion Monte Carlo}},}\ }\href {\doibase
  10.1103/PhysRevC.67.065806} {\bibfield  {journal} {\bibinfo  {journal} {Phys.
  Rev. C}\ }\textbf {\bibinfo {volume} {67}},\ \bibinfo {pages} {065806}
  (\bibinfo {year} {2003})},\ \Eprint {http://arxiv.org/abs/nucl-th/0304042}
  {arXiv:nucl-th/0304042} \BibitemShut {NoStop}%
\bibitem [{\citenamefont {Schiavilla}\ \emph {et~al.}(2021)\citenamefont
  {Schiavilla}, \citenamefont {Girlanda}, \citenamefont {Gnech}, \citenamefont
  {Kievsky}, \citenamefont {Lovato}, \citenamefont {Marcucci}, \citenamefont
  {Piarulli},\ and\ \citenamefont {Viviani}}]{Schiavilla:2021dun}%
  \BibitemOpen
  \bibfield  {author} {\bibinfo {author} {\bibfnamefont {R.}~\bibnamefont
  {Schiavilla}}, \bibinfo {author} {\bibfnamefont {L.}~\bibnamefont
  {Girlanda}}, \bibinfo {author} {\bibfnamefont {A.}~\bibnamefont {Gnech}},
  \bibinfo {author} {\bibfnamefont {A.}~\bibnamefont {Kievsky}}, \bibinfo
  {author} {\bibfnamefont {A.}~\bibnamefont {Lovato}}, \bibinfo {author}
  {\bibfnamefont {L.~E.}\ \bibnamefont {Marcucci}}, \bibinfo {author}
  {\bibfnamefont {M.}~\bibnamefont {Piarulli}}, \ and\ \bibinfo {author}
  {\bibfnamefont {M.}~\bibnamefont {Viviani}},\ }\bibfield  {title} {\enquote
  {\bibinfo {title} {{Two- and three-nucleon contact interactions and
  ground-state energies of light- and medium-mass nuclei}},}\ }\href {\doibase
  10.1103/PhysRevC.103.054003} {\bibfield  {journal} {\bibinfo  {journal}
  {Phys. Rev. C}\ }\textbf {\bibinfo {volume} {103}},\ \bibinfo {pages}
  {054003} (\bibinfo {year} {2021})},\ \Eprint
  {http://arxiv.org/abs/2102.02327} {arXiv:2102.02327 [nucl-th]} \BibitemShut
  {NoStop}%
\bibitem [{\citenamefont {K\"onig}\ \emph {et~al.}(2017)\citenamefont
  {K\"onig}, \citenamefont {Grie\ss{}hammer}, \citenamefont {Hammer},\ and\
  \citenamefont {van Kolck}}]{Konig:2016utl}%
  \BibitemOpen
  \bibfield  {author} {\bibinfo {author} {\bibfnamefont {Sebastian}\
  \bibnamefont {K\"onig}}, \bibinfo {author} {\bibfnamefont {Harald~W.}\
  \bibnamefont {Grie\ss{}hammer}}, \bibinfo {author} {\bibfnamefont {H.~W.}\
  \bibnamefont {Hammer}}, \ and\ \bibinfo {author} {\bibfnamefont
  {U.}~\bibnamefont {van Kolck}},\ }\bibfield  {title} {\enquote {\bibinfo
  {title} {{Nuclear Physics Around the Unitarity Limit}},}\ }\href {\doibase
  10.1103/PhysRevLett.118.202501} {\bibfield  {journal} {\bibinfo  {journal}
  {Phys. Rev. Lett.}\ }\textbf {\bibinfo {volume} {118}},\ \bibinfo {pages}
  {202501} (\bibinfo {year} {2017})},\ \Eprint
  {http://arxiv.org/abs/1607.04623} {arXiv:1607.04623 [nucl-th]} \BibitemShut
  {NoStop}%
\bibitem [{\citenamefont {Kievsky}\ \emph {et~al.}(2018)\citenamefont
  {Kievsky}, \citenamefont {Viviani}, \citenamefont {Logoteta}, \citenamefont
  {Bombaci},\ and\ \citenamefont {Girlanda}}]{Kievsky:2018xsl}%
  \BibitemOpen
  \bibfield  {author} {\bibinfo {author} {\bibfnamefont {A.}~\bibnamefont
  {Kievsky}}, \bibinfo {author} {\bibfnamefont {M.}~\bibnamefont {Viviani}},
  \bibinfo {author} {\bibfnamefont {D.}~\bibnamefont {Logoteta}}, \bibinfo
  {author} {\bibfnamefont {I.}~\bibnamefont {Bombaci}}, \ and\ \bibinfo
  {author} {\bibfnamefont {L.}~\bibnamefont {Girlanda}},\ }\bibfield  {title}
  {\enquote {\bibinfo {title} {{Correlations imposed by the unitary limit
  between few-nucleon systems, nuclear matter and neutron stars}},}\ }\href
  {\doibase 10.1103/PhysRevLett.121.072701} {\bibfield  {journal} {\bibinfo
  {journal} {Phys. Rev. Lett.}\ }\textbf {\bibinfo {volume} {121}},\ \bibinfo
  {pages} {072701} (\bibinfo {year} {2018})},\ \Eprint
  {http://arxiv.org/abs/1806.02636} {arXiv:1806.02636 [nucl-th]} \BibitemShut
  {NoStop}%
\bibitem [{\citenamefont {Wiringa}\ \emph {et~al.}(1995)\citenamefont
  {Wiringa}, \citenamefont {Stoks},\ and\ \citenamefont
  {Schiavilla}}]{Wiringa:1994wb}%
  \BibitemOpen
  \bibfield  {author} {\bibinfo {author} {\bibfnamefont {Robert~B.}\
  \bibnamefont {Wiringa}}, \bibinfo {author} {\bibfnamefont {V.~G.~J.}\
  \bibnamefont {Stoks}}, \ and\ \bibinfo {author} {\bibfnamefont
  {R.}~\bibnamefont {Schiavilla}},\ }\bibfield  {title} {\enquote {\bibinfo
  {title} {{An Accurate nucleon-nucleon potential with charge independence
  breaking}},}\ }\href {\doibase 10.1103/PhysRevC.51.38} {\bibfield  {journal}
  {\bibinfo  {journal} {Phys. Rev. C}\ }\textbf {\bibinfo {volume} {51}},\
  \bibinfo {pages} {38--51} (\bibinfo {year} {1995})},\ \Eprint
  {http://arxiv.org/abs/nucl-th/9408016} {arXiv:nucl-th/9408016} \BibitemShut
  {NoStop}%
\bibitem [{\citenamefont {Pudliner}\ \emph {et~al.}(1995)\citenamefont
  {Pudliner}, \citenamefont {Pandharipande}, \citenamefont {Carlson},\ and\
  \citenamefont {Wiringa}}]{Pudliner:1995wk}%
  \BibitemOpen
  \bibfield  {author} {\bibinfo {author} {\bibfnamefont {B.~S.}\ \bibnamefont
  {Pudliner}}, \bibinfo {author} {\bibfnamefont {V.~R.}\ \bibnamefont
  {Pandharipande}}, \bibinfo {author} {\bibfnamefont {J.}~\bibnamefont
  {Carlson}}, \ and\ \bibinfo {author} {\bibfnamefont {Robert~B.}\ \bibnamefont
  {Wiringa}},\ }\bibfield  {title} {\enquote {\bibinfo {title} {{Quantum Monte
  Carlo calculations of A \ensuremath{<}= 6 nuclei}},}\ }\href {\doibase
  10.1103/PhysRevLett.74.4396} {\bibfield  {journal} {\bibinfo  {journal}
  {Phys. Rev. Lett.}\ }\textbf {\bibinfo {volume} {74}},\ \bibinfo {pages}
  {4396--4399} (\bibinfo {year} {1995})},\ \Eprint
  {http://arxiv.org/abs/nucl-th/9502031} {arXiv:nucl-th/9502031} \BibitemShut
  {NoStop}%
\bibitem [{\citenamefont {Akmal}\ \emph {et~al.}(1998)\citenamefont {Akmal},
  \citenamefont {Pandharipande},\ and\ \citenamefont
  {Ravenhall}}]{Akmal:1998cf}%
  \BibitemOpen
  \bibfield  {author} {\bibinfo {author} {\bibfnamefont {A.}~\bibnamefont
  {Akmal}}, \bibinfo {author} {\bibfnamefont {V.~R.}\ \bibnamefont
  {Pandharipande}}, \ and\ \bibinfo {author} {\bibfnamefont {D.~G.}\
  \bibnamefont {Ravenhall}},\ }\bibfield  {title} {\enquote {\bibinfo {title}
  {{The Equation of state of nucleon matter and neutron star structure}},}\
  }\href {\doibase 10.1103/PhysRevC.58.1804} {\bibfield  {journal} {\bibinfo
  {journal} {Phys. Rev. C}\ }\textbf {\bibinfo {volume} {58}},\ \bibinfo
  {pages} {1804--1828} (\bibinfo {year} {1998})},\ \Eprint
  {http://arxiv.org/abs/nucl-th/9804027} {arXiv:nucl-th/9804027} \BibitemShut
  {NoStop}%
\bibitem [{\citenamefont {Machleidt}\ and\ \citenamefont
  {Entem}(2011)}]{Machleidt2011}%
  \BibitemOpen
  \bibfield  {author} {\bibinfo {author} {\bibfnamefont {R.}~\bibnamefont
  {Machleidt}}\ and\ \bibinfo {author} {\bibfnamefont {D.R.}\ \bibnamefont
  {Entem}},\ }\bibfield  {title} {\enquote {\bibinfo {title} {Chiral effective
  field theory and nuclear forces},}\ }\href {\doibase
  https://doi.org/10.1016/j.physrep.2011.02.001} {\bibfield  {journal}
  {\bibinfo  {journal} {Phys. Rep.}\ }\textbf {\bibinfo {volume} {503}},\
  \bibinfo {pages} {1} (\bibinfo {year} {2011})}\BibitemShut {NoStop}%
\bibitem [{\citenamefont {Moreno}\ \emph {et~al.}(2022)\citenamefont {Moreno},
  \citenamefont {Carleo}, \citenamefont {Georges},\ and\ \citenamefont
  {Stokes}}]{Moreno2022}%
  \BibitemOpen
  \bibfield  {author} {\bibinfo {author} {\bibfnamefont {Javier~Robledo}\
  \bibnamefont {Moreno}}, \bibinfo {author} {\bibfnamefont {Giuseppe}\
  \bibnamefont {Carleo}}, \bibinfo {author} {\bibfnamefont {Antoine}\
  \bibnamefont {Georges}}, \ and\ \bibinfo {author} {\bibfnamefont {James}\
  \bibnamefont {Stokes}},\ }\bibfield  {title} {\enquote {\bibinfo {title}
  {Fermionic wave functions from neural-network constrained hidden states},}\
  }\href {\doibase 10.1073/PNAS.2122059119/SUPPL_FILE/PNAS.2122059119.SAPP.PDF}
  {\bibfield  {journal} {\bibinfo  {journal} {Proceedings of the National
  Academy of Sciences of the United States of America}\ }\textbf {\bibinfo
  {volume} {119}},\ \bibinfo {pages} {e2122059119} (\bibinfo {year}
  {2022})}\BibitemShut {NoStop}%
\bibitem [{\citenamefont {{Zaheer}}\ \emph {et~al.}(2017)\citenamefont
  {{Zaheer}}, \citenamefont {{Kottur}}, \citenamefont {{Ravanbakhsh}},
  \citenamefont {{Poczos}}, \citenamefont {{Salakhutdinov}},\ and\
  \citenamefont {{Smola}}}]{Zaheer:2017}%
  \BibitemOpen
  \bibfield  {author} {\bibinfo {author} {\bibfnamefont {Manzil}\ \bibnamefont
  {{Zaheer}}}, \bibinfo {author} {\bibfnamefont {Satwik}\ \bibnamefont
  {{Kottur}}}, \bibinfo {author} {\bibfnamefont {Siamak}\ \bibnamefont
  {{Ravanbakhsh}}}, \bibinfo {author} {\bibfnamefont {Barnabas}\ \bibnamefont
  {{Poczos}}}, \bibinfo {author} {\bibfnamefont {Ruslan}\ \bibnamefont
  {{Salakhutdinov}}}, \ and\ \bibinfo {author} {\bibfnamefont {Alexander}\
  \bibnamefont {{Smola}}},\ }\bibfield  {title} {\enquote {\bibinfo {title}
  {{Deep Sets}},}\ }\href@noop {} {\bibfield  {journal} {\bibinfo  {journal}
  {arXiv e-prints}\ ,\ \bibinfo {eid} {arXiv:1703.06114}} (\bibinfo {year}
  {2017})},\ \Eprint {http://arxiv.org/abs/1703.06114} {arXiv:1703.06114
  [cs.LG]} \BibitemShut {NoStop}%
\bibitem [{\citenamefont {{Wagstaff}}\ \emph {et~al.}(2019)\citenamefont
  {{Wagstaff}}, \citenamefont {{Fuchs}}, \citenamefont {{Engelcke}},
  \citenamefont {{Posner}},\ and\ \citenamefont {{Osborne}}}]{Wagstaff:2019}%
  \BibitemOpen
  \bibfield  {author} {\bibinfo {author} {\bibfnamefont {Edward}\ \bibnamefont
  {{Wagstaff}}}, \bibinfo {author} {\bibfnamefont {Fabian~B.}\ \bibnamefont
  {{Fuchs}}}, \bibinfo {author} {\bibfnamefont {Martin}\ \bibnamefont
  {{Engelcke}}}, \bibinfo {author} {\bibfnamefont {Ingmar}\ \bibnamefont
  {{Posner}}}, \ and\ \bibinfo {author} {\bibfnamefont {Michael}\ \bibnamefont
  {{Osborne}}},\ }\bibfield  {title} {\enquote {\bibinfo {title} {{On the
  Limitations of Representing Functions on Sets}},}\ }\href@noop {} {\bibfield
  {journal} {\bibinfo  {journal} {arXiv e-prints}\ ,\ \bibinfo {eid}
  {arXiv:1901.09006}} (\bibinfo {year} {2019})},\ \Eprint
  {http://arxiv.org/abs/1901.09006} {arXiv:1901.09006 [cs.LG]} \BibitemShut
  {NoStop}%
\bibitem [{\citenamefont {Pescia}\ \emph {et~al.}(2022)\citenamefont {Pescia},
  \citenamefont {Han}, \citenamefont {Lovato}, \citenamefont {Lu},\ and\
  \citenamefont {Carleo}}]{Pescia:2021kxb}%
  \BibitemOpen
  \bibfield  {author} {\bibinfo {author} {\bibfnamefont {Gabriel}\ \bibnamefont
  {Pescia}}, \bibinfo {author} {\bibfnamefont {Jiequn}\ \bibnamefont {Han}},
  \bibinfo {author} {\bibfnamefont {Alessandro}\ \bibnamefont {Lovato}},
  \bibinfo {author} {\bibfnamefont {Jianfeng}\ \bibnamefont {Lu}}, \ and\
  \bibinfo {author} {\bibfnamefont {Giuseppe}\ \bibnamefont {Carleo}},\
  }\bibfield  {title} {\enquote {\bibinfo {title} {{Neural-network quantum
  states for periodic systems in continuous space}},}\ }\href {\doibase
  10.1103/PhysRevResearch.4.023138} {\bibfield  {journal} {\bibinfo  {journal}
  {Phys. Rev. Res.}\ }\textbf {\bibinfo {volume} {4}},\ \bibinfo {pages}
  {023138} (\bibinfo {year} {2022})},\ \Eprint
  {http://arxiv.org/abs/2112.11957} {arXiv:2112.11957 [quant-ph]} \BibitemShut
  {NoStop}%
\bibitem [{\citenamefont {Sarsa}\ \emph {et~al.}(2003)\citenamefont {Sarsa},
  \citenamefont {Fantoni}, \citenamefont {Schmidt},\ and\ \citenamefont
  {Pederiva}}]{Sarsa:2003zu}%
  \BibitemOpen
  \bibfield  {author} {\bibinfo {author} {\bibfnamefont {A.}~\bibnamefont
  {Sarsa}}, \bibinfo {author} {\bibfnamefont {S.}~\bibnamefont {Fantoni}},
  \bibinfo {author} {\bibfnamefont {K.~E.}\ \bibnamefont {Schmidt}}, \ and\
  \bibinfo {author} {\bibfnamefont {F.}~\bibnamefont {Pederiva}},\ }\bibfield
  {title} {\enquote {\bibinfo {title} {{Neutron matter at zero temperature with
  auxiliary field diffusion Monte Carlo}},}\ }\href {\doibase
  10.1103/PhysRevC.68.024308} {\bibfield  {journal} {\bibinfo  {journal} {Phys.
  Rev.}\ }\textbf {\bibinfo {volume} {C68}},\ \bibinfo {pages} {024308}
  (\bibinfo {year} {2003})},\ \Eprint {http://arxiv.org/abs/nucl-th/0303035}
  {arXiv:nucl-th/0303035 [nucl-th]} \BibitemShut {NoStop}%
\bibitem [{\citenamefont {{Metropolis}}\ \emph {et~al.}(1953)\citenamefont
  {{Metropolis}}, \citenamefont {{Rosenbluth}}, \citenamefont {{Rosenbluth}},
  \citenamefont {{Teller}},\ and\ \citenamefont {{Teller}}}]{Metropolis:1953}%
  \BibitemOpen
  \bibfield  {author} {\bibinfo {author} {\bibfnamefont {Nicholas}\
  \bibnamefont {{Metropolis}}}, \bibinfo {author} {\bibfnamefont {Arianna~W.}\
  \bibnamefont {{Rosenbluth}}}, \bibinfo {author} {\bibfnamefont {Marshall~N.}\
  \bibnamefont {{Rosenbluth}}}, \bibinfo {author} {\bibfnamefont {Augusta~H.}\
  \bibnamefont {{Teller}}}, \ and\ \bibinfo {author} {\bibfnamefont {Edward}\
  \bibnamefont {{Teller}}},\ }\bibfield  {title} {\enquote {\bibinfo {title}
  {{Equation of State Calculations by Fast Computing Machines}},}\ }\href
  {\doibase 10.1063/1.1699114} {\bibfield  {journal} {\bibinfo  {journal}
  {\jcp}\ }\textbf {\bibinfo {volume} {21}},\ \bibinfo {pages} {1087--1092}
  (\bibinfo {year} {1953})}\BibitemShut {NoStop}%
\bibitem [{\citenamefont {Gandolfi}\ \emph {et~al.}(2009)\citenamefont
  {Gandolfi}, \citenamefont {Illarionov}, \citenamefont {Pederiva},
  \citenamefont {Schmidt},\ and\ \citenamefont {Fantoni}}]{Gandolfi2009}%
  \BibitemOpen
  \bibfield  {author} {\bibinfo {author} {\bibfnamefont {S.}~\bibnamefont
  {Gandolfi}}, \bibinfo {author} {\bibfnamefont {A.~Yu}\ \bibnamefont
  {Illarionov}}, \bibinfo {author} {\bibfnamefont {F.}~\bibnamefont
  {Pederiva}}, \bibinfo {author} {\bibfnamefont {K.~E.}\ \bibnamefont
  {Schmidt}}, \ and\ \bibinfo {author} {\bibfnamefont {S.}~\bibnamefont
  {Fantoni}},\ }\bibfield  {title} {\enquote {\bibinfo {title} {Equation of
  state of low-density neutron matter, and the 1s0 pairing gap},}\ }\href
  {\doibase 10.1103/PHYSREVC.80.045802/FIGURES/6/MEDIUM} {\bibfield  {journal}
  {\bibinfo  {journal} {Physical Review C - Nuclear Physics}\ }\textbf
  {\bibinfo {volume} {80}},\ \bibinfo {pages} {045802} (\bibinfo {year}
  {2009})}\BibitemShut {NoStop}%
\bibitem [{\citenamefont {Piarulli}\ \emph {et~al.}(2018)\citenamefont
  {Piarulli} \emph {et~al.}}]{Piarulli:2017dwd}%
  \BibitemOpen
  \bibfield  {author} {\bibinfo {author} {\bibfnamefont {M.}~\bibnamefont
  {Piarulli}} \emph {et~al.},\ }\bibfield  {title} {\enquote {\bibinfo {title}
  {{Light-nuclei spectra from chiral dynamics}},}\ }\href {\doibase
  10.1103/PhysRevLett.120.052503} {\bibfield  {journal} {\bibinfo  {journal}
  {Phys. Rev. Lett.}\ }\textbf {\bibinfo {volume} {120}},\ \bibinfo {pages}
  {052503} (\bibinfo {year} {2018})},\ \Eprint
  {http://arxiv.org/abs/1707.02883} {arXiv:1707.02883 [nucl-th]} \BibitemShut
  {NoStop}%
\bibitem [{\citenamefont {Negele}\ and\ \citenamefont
  {Vautherin}(1973)}]{Negele:1971vb}%
  \BibitemOpen
  \bibfield  {author} {\bibinfo {author} {\bibfnamefont {John~W.}\ \bibnamefont
  {Negele}}\ and\ \bibinfo {author} {\bibfnamefont {D.}~\bibnamefont
  {Vautherin}},\ }\bibfield  {title} {\enquote {\bibinfo {title} {{Neutron star
  matter at subnuclear densities}},}\ }\href {\doibase
  10.1016/0375-9474(73)90349-7} {\bibfield  {journal} {\bibinfo  {journal}
  {Nucl. Phys. A}\ }\textbf {\bibinfo {volume} {207}},\ \bibinfo {pages}
  {298--320} (\bibinfo {year} {1973})}\BibitemShut {NoStop}%
\end{thebibliography}%

\end{document}